\documentclass[preprint2]{aastex}

\shortauthors{KIM, SUNG, \& LEE}
\shorttitle{NGC 300}

\begin{document}

\title{Stellar Contents and Globular Cluster Candidates 
  in the Sculptor Group Galaxy NGC 300}
\author{Sang Chul Kim, Hwankyung Sung, and Myung Gyoon Lee}

\affil{Astronomy Program, SEES, Seoul National University, Seoul 151-742, Korea}
\affil{\it E-mail: sckim@astro.snu.ac.kr, sungh@astro.snu.ac.kr, mglee@astrog.snu.ac.kr}
 
\vskip 1cm
\affil{[Accepted for publication in the Journal of the Korean Astronomical Society, 2002 March issue]}

\begin{abstract}
We present $UBVI$ CCD photometry of the stellar contents and
globular cluster(GC) candidates in the spiral galaxy NGC 300 in the Sculptor group.
Color-magnitude diagrams 
for 18 OB associations having more than 30 member stars are presented.
The slope of the initial mass function 
for the bright stars in NGC 300 is estimated to be 
$\Gamma = -2.6 \pm 0.3$.  
Assuming the distance to NGC 300 of $(m-M)_0 = 26.53\pm 0.07$,
the mean absolute magnitude of three brightest
blue stars is obtained to be $<M_V^{BSG}(3)>$ = $-8.95$ mag.
We have performed search for GCs in NGC 300 and have found 
17 GC candidates in this galaxy. 
Some characteristics of these GC candidates are discussed.
\end{abstract}
 
 
\section{INTRODUCTION}

Sculptor group is the nearest galaxy group to the Local Group 
containing five major spiral galaxies 
(NGC 55, NGC 247, NGC 253, NGC 300, and NGC 7793) and 
$\sim$20 dwarf galaxies
(C{\^o}t{\'e} et al. 1997, Whiting 1999; cf. van den Bergh 1999).
Due to the rather small distance to the Sculptor group,
there have been many studies on the galaxies in this group.

NGC 300 (=ESO 295-20, IRAS 00523-3756, PGC 3238) is 
an SA(s)d spiral galaxy in the Sculptor group.
Basic information of this galaxy is summarized in Table 1.
It is a rather bright ($M_B = -18.6$) and nearly face-on galaxy,
similar to M33 seen in the northern hemisphere (Blair \& Long 1997).
There have been several studies on NGC 300, and some of them are summarized 
in the following.
Pietrzy\'nski et al. (2001) have found 117 OB associations in NGC 300,
Davidge (1998) has studied the evolved red stellar contents from $JHK$ photometry, 
Breysacher et al. (1997) have studied Wolf-Rayet stars detected in five associations of NGC 300,
and Soffner et al. (1996) have searched for and found 34 planetary nebulae and 
88 {\sc{H~ii}} regions.
Freedman et al. (1992) obtained the distance modulus to this galaxy of 
   $(m-M)_0 = 26.66\pm 0.10$ mag (d = $2.1 \pm 0.1$ Mpc) using 16 known Cepheid variables.
Pannuti et al. (2000) have conducted supernova remnant (SNR) search in NGC 300
using X-ray, optical, and radio observations,
resulting in total 44 SNR candidates including 16 new objects. 
Recently Read \& Pietsch (2001) have studied X-ray emission sources in NGC 300 
using $ROSAT$ data.
Freedman et al. (2001) have revised the distance modulus to NGC 300 
using the {\it Hubble Space Telescope} Key Project Cepheid data to be 
$(m-M)_0 = 26.53\pm 0.07$ mag (d = $2.02 \pm 0.07$ Mpc),
which is adopted in this study.

  There have been a few surveys 
for globular clusters(GCs) in the Sculptor group galaxies.
From excellent seeing plates and spectrophotometry, Da Costa \& Graham (1982) have
found three GCs in edge-on galaxy NGC 55.
Liller \& Alcaino (1983a) have found 51 GC candidates in NGC 55
from excellent-seeing plates taken with ESO 3.6m telescope.
Liller \& Alcaino (1983b) also have found 63 GC candidates in 
NGC 253 from excellent-seeing plates taken with ESO 3.6m telescope.
In NGC 253, Blecha (1986) have found 25 GC candidates 
from using high-resolution-wide field McMullan electronographic camera attached
to the Danish 1.5m telescope. 
Later, for the 57 and 58 GC candidates in NGC 55 and NGC 253, respectivey,
Beasley \& Sharples (2000) have performed spectroscopic GC survey and
identified one probable GC and 14 GCs in NGC 55 and NGC 253, respectivey.
And they have estimated total number of GCs in NGC 55 and NGC 253
to be $\sim$20 and $\sim$60, respectivey.
Watson et al. (1996) have discovered four young, luminous, compact stellar clusters
in the central region of the starburst spiral galaxy NGC 253, which are believed
to be excellent candidates of young GCs.
Recently Jerjen \& Rejkuba (2001) have found one GC candidate 
in dwarf galaxy ESO 540-032 using Very Large Telescope UT1+FORS1.
However, little is known about GCs in NGC 300.  
Only Bright \& Olsen (2000) have
indicated that they obtained 100 -- 150 GC candidates in each of seven brightest
galaxies in the Sculptor group.  But the detailed information is not yet published.
In this paper we report the first list of GC candidates in NGC 300.

This paper is composed as follows. 
Section II describes observations and data reduction, and
Section III presents photometric diagrams and photometric analyses
of 18 OB associations in NGC 300.
Section IV presents the search method and the result for new GC candidates, and
Section V discusses mass function of bright stars, brightest stars,
some characteristics of GCs, and the nucleus of NGC 300.
Finally, a summary and conclusions are given in Section VI.

\section{OBSERVATIONS}

$UBVI$ CCD photometry was performed on 1997 November 23 at 
the Siding Spring Observatory with the 40 inch (1m) telescope (f/8) 
and a thinned
SITe 2048 $\times$ 2048 CCD (24$\mu$m pixels).  The scale was 0.$''$602
pixel$^{-1}$, giving 20.$'$5 on a side. Exposure times used in the observations
were 60 s and 120 s in $I$, 300 s in $V$, 600 s in $B$, 1200 s in $U$.
The night was photometric and the seeing was 1.$''$4 in the 300 sec $V$ image.
A finding chart for the observed area is shown in Fig.~1. In the figure
we marked only stars with $V$ = 16.5 -- 21 mag to show the spatial distribution
of bright stars in NGC 300. In addition we also marked
18 rich OB associations and 17 GC candidates.

All the preprocessing, such as overscan correction, bias subtraction and flat
fielding were done using the IRAF\footnote{IRAF is distributed by the National
Optical Astronomy Observatories, which are operated by the Association of
Universities for Research in Astronomy, Inc.}/{\small CCDRED} package.
Instrumental magnitudes were obtained using the IRAF version of DAOPHOT II
(Stetson 1990) via point spread function (PSF) fitting. All the instrumental
magnitudes were transformed to the standard Johnson-Cousins $UBVI$ system
using SAAO observations of equatorial standards (Menzies et al. 1991) and blue and
red standards by Kilkenny et al. (1998). The primary extinction coefficients were
determined and used.
Details involving standard transformations can be found in Sung \& Bessell (2000).

The total number of stars which were measured down to $V$ = 22 mag is 6027.
Table 2 lists the photometry of 167 stars with $V < 19.0$ mag and
contains the running number, $\alpha$ (J2000.0), $\delta$ (J2000.0),
the weighted mean values of the magnitude and color, the weighted errors, and
the number of independent measurements.
The variable stars found by Graham (1984), 
the photometric sequence of Graham (1981) and Walker (1995), and
GC candidates newly found in this paper are identified
in the last two columns.

The mean values of the errors are listed in Table 3 (``N'' in Table 3 represents
the number of stars used in the statistics). The $UBVI$ magnitude and
colors for $V_{\rm CCD} \leq 20$ mag are compared with previous photoelectric
or CCD photometry by Graham (1981) and Walker (1995) in Fig.~2, showing differences in
the sense, previous photometry minus our new photometry. Our photometry
is 
consistent with that by Graham (1981) or by Walker (1995) in $B-V$ and $V-I$
($\Delta (B-V) = +0.008 \pm 0.013$ and $\Delta (V-I) = -0.007 \pm 0.014$), but slightly
fainter in $V$ ($\Delta V = -0.025 \pm 0.024$) and redder in $U-B$ ($\Delta (U-B)
= -0.042 \pm 0.016$).

\section{PHOTOMETRIC DIAGRAMS AND OB ASSOCIATIONS}

\subsection{Photometric Diagrams}

As NGC 300 is a member of Sculptor group of galaxies ($d \approx$ 2.02 Mpc
: Freedman et al. 2001),
the stars in the galaxy are very faint. In addition many bright blue stars
are members of OB associations, and therefore their photometric values contain
inevitably large errors due to crowding effect even though we paid much
attention to the photometry. 
The resulting color-magnitude diagrams (CMDs)
from our photometry are presented in Fig.~3. 
In the figure we marked the GC candidates (open circle
with dot; see Section IV for the selection criteria) and
variable stars found by Graham (1984) (open square with dot).
Among 34 variable stars, 22 were identified from the coordinates
and charts given by Graham (1984) (The coordiantes given by him are not
accurate enough to identify the variables unambiguously). 
Among the remaining 12 stars, 9 stars are outside our field of view, 
3 stars are too faint to be measured from our images.

The ($V, V-I$) diagram (Fig.~3 (a)) do not give much information
on the photometric characteristics of the stars in NGC 300 due partly
to the short exposures of $I$ images and partly to the characteristic
nature of $V-I$ color, which is insensitive to the temperature variation of
blue stars. On the other hand, $U-B$ is rather sensitive to the temperature
variation of hot stars, but the blue stars in NGC 300 are very faint in
$U$ pass band.  Therefore the ($V, U-B$) CMD also does not give
much information on the faint stars. 
Among the three CMDs, the ($V, B-V$) diagram
shows well-defined blue stars more clearly. 
We will use only the ($V, B-V$) diagram in the analysis of
the photometric characteristics of stars in NGC 300.
In this figure, a large number of faint blue stars can easily be found. 
They are bona fide members of NGC 300.
Apparently stars brighter than $V$ = 17.5 mag seem to be
foreground thick disk or halo stars in our Galaxy.

We presented the ($U-B, B-V$) diagram in Fig.~4. As the most
bright blue stars are suspected to be blue supergiants in NGC 300,
we drew the intrinsic relation for supergiants (Sung 1995) in the figure. 
Even though the photometric error in $B-V$ as well as $U-B$ is large,
most blue stars are scattered around the intrinsic relation for
supergiants, indicating that the internal reddening
is very small. There are several yellow stars ($B-V$ = 0.4 -- 0.8 mag).
A few of them are GC candidates, but
the remainings are foreground stars in the Galaxy. 
As they are suspected as thick disk or halo stars in the Galaxy, 
they show evident UV excesses. 
The others with abnormal red colors in the diagram are faint red stars. 
Many of them ($V <$ 17.5 mag) are foreground red stars in the galaxy, but a few
faint red stars may be red supergiant stars in NGC 300.
Their abnormal colors are mainly due to large photometric errors.
As NGC 300 is located near the south galactic pole ($b = -79.\arcdeg42$),
the foreground reddening due to the interstellar cloud in the Galaxy is practically
zero. In addition NGC 300 is nearly face-on, and therefore the internal
reddening may not be very large. 
For the reddening correction of stars in NGC 300
we have used the value $E(B-V)$ = 0.013 mag obtained by 
Schlegel, Finkbeiner, \& Davis (1998).

\subsection{OB Associations}

Recently Pietrzy\'nski et al. (2001) published a catalog of OB associations in NGC 300.
They found 117 OB associations using Path Linkage Criterion technique.
The telescope they used is the ESO/MPI 2.2 m, and therefore the limiting
magnitude of their photometry is about 1 mag fainter than ours. 
To derive the age distribution of OB associations in NGC 300, 
the OB associations having more than 10 stars in our photometry
were selected.
In the analysis, we limited our attention to the OB associations
having more than 30 member stars in the catalog ($N_* > 30$).
Among 117 OB associations, 18 OB associations are satisfied the above criterion.
We present 18 CMDs with appropriate isochrones of the Padova group 
(Bertelli et al. 1994) in Fig.~5. 
Due to the large photometric errors of faint stars
in NGC 300, the CMD of individual OB association shows a large scatter.
The color of faint stars are more affected by crowding
effect (see Sung et al. (1999) for the effect of crowding on color). The large
scatter in $B-V$ is therefore due partly to the crowding effect and partly 
to the contamination of faint field stars in NGC 300. 
In addition we cannot
exclude the possibility of the contamination of foreground stars in the Galaxy
or GCs in NGC 300 among yellow or red stars in the line of
sight. We determined the age of each association by paying more attention
to the bright blue stars in the association.

To obtain the reliable ages of OB associations or young open clusters,
spectral type and luminosity class should be known because the photometric
indices such as $B-V$ or even $U-B$ are inadequate for the accurate
estimate of temperature of O- and early B-type stars. Actually no
information on the spectral type of blue stars in NGC 300 is available
currently. 
The ages of OB associations in NGC 300 have, therefore,
inevitably large errors if we determine the age from photometric colors alone. 
We should determine
the age of each OB association with such a limitation.
Among the 18 OB associations, 
13 associations have ages younger than 10 Myr ($\log \tau < 7$).
Many of them may have much younger ages (only about a few Myr) because there are
a few very bright blue stars ($M_V < -8$ mag). 
The associations that belong to this category are 
AS\_016, \_029, \_052, \_082, \_084, \_086, and \_102. For
these associations the ages from the CMDs are highly uncertain.
There are five associations (AS\_018, \_026, \_027, \_058, and \_096) 
with ages older than 10 Myr. 
They have apparently no O-type stars and seem to be old
B associations. A few OB associations seem to contain a red supergiant
candidate (AS\_002, \_009, and less luminous red supergiant
candidates in AS\_101).

Spatially these rich OB associations ($N_* > 30$) are more concentrated in
the NW part of the galaxy, but the brightest OB association in NGC 300
is AS\_102, about 4$'$ E of the center of NGC 300. 
Two out of three brightest blue
stars in NGC 300 are in this association. In the H$\alpha$ image of NGC 300,
AS\_102 is the brightest in H$\alpha$, which indicates that AS\_102 is
associated with the brightest {\sc{H~ii}} region. AS\_102 is the most active
star forming region in NGC 300. There are many bright blue stars in the CMD
of AS\_102. Even though the photometric error is large, 
there is few red supergiant candiates in the association. 
This fact again implies that the association is extremely young. 
The second and third brightest OB associations are AS\_052 and
\_084. Even though AS\_052 is the richest OB association ($N_* = 101$) in NGC 300, 
the brightest star in the associaton is about 0.8 mag fainter that the brightest
star in AS\_102.

\section{GLOBULAR CLUSTER CANDIDATES}

\subsection{Search Method}

Globular clusters in NGC 300 are barely resolved in our images.
We have used photometric information, morphological parameters and visual
inspection to survey globular cluster candidates as follows.
 From the ($V, B-V$) CMD in Fig. 3, we have first 
taken the point sources with 
$V < 20.0$ mag and $0.3 < (B-V) < 2.0$ to search for globular cluster candidates.  
We have considered $V = 20.0$ mag as the reliable magnitude limit
and $0.3 < (B-V) < 2.0$ to be the safe color range that contains
all the GC candidates in NGC 300.
For these selected point sources
we have calculated the morphological image classification parameters
($r_1$ and $r_{-2}$ radial moments) in three ($V$-band and two $I$-band) 
best seeing images.
The two radial moments are effective image radii in pixel units,
where $r_1$ is an indicator of image wing spread and
$r_{-2}$ indicates the image central concentration
(see Harris et al. 1991; Kim et al. 2002; Lee et al. 2002).
  Because the images have degradations at the corners, we only considered
the objects inside of a circle centered at the image center and 
with the diameter of the image width ($20.'5$).

  Fig. 6 shows the $r_1$ and $r_{-2}$ parameters calculated 
on the $V$-band image.  
The horizontal sequence of objects are mainly stars, 
objects with a little larger moment values should be GCs in NGC 300,
and some objects with largest moment values are probably background galaxies.
The box region contains the objects for which we have performed 
visual examination on each image and investigated radial, contour, 
and surface profiles.
The circled objects denote the resulting GC candidates and
the size of the circle is proportional to the probability to be a GC;
the largest circles denote probable GC candidates (class 1),
the intermediate size circles denote the possible candidates (class 2),
and the smallest circles denote doubtful candidates (class 3).

Fig. 7 shows the $V$-band mosaic images and radial profiles 
of a typical star and a class 1 GC candidate discovered in this study.
The FWHM of the star in Fig. 7 is $\sim 2.4$ pixel, while 
that of the GC candidate is $\sim$ 2.8 pixel.
Though the GC candidate is fainter than the star ($V$=18.2 versus $V$=17.4),
it is clear that the GC candidate is more extended than the star.

We also have performed smoothing the images 
using the IRAF/{\small RMEDIAN} package
and subtracted the results from the originals to form the final images 
on which the GC search was made again.
While these images clearly show the fine structures of NGC 300 
especially in the nuclear region and patchy regions,
they are not much better in the search for GCs
than the original images.

There are several sources that we could not be able to 
investigate the image structure due to the crowding of 
two or more objects in small region and large pixel size 
of our images (1 px $=$ $0.''602$ = 5.89 pc assuming d = 2.02 
Mpc, and so $1''$ $=$ 9.79 pc).
Imaing data of better quality (e.g. Bright \& Olsen 2000) will
be helpful to uncover the nature of these objects.

\subsection{Search Result}

In Table 4 we present the newly discovered GC candidates in NGC 300.
Column 1 gives the GC candidate ID numbers.
Columns 2 and 3 are the RA and DEC coordinates (epoch J2000), 
columns 4-7 are $V$ magnitude,
and $V-I$, $B-V$, and $U-B$ colors, respectively, 
columns 8-11 are the errors in $V$, 
$V-I$, $B-V$ and $U-B$, respectively, 
columns 12-15 are the number of independent measurements
and the last column is the class of the candidates.

The positions of these GC candidates in the CMD and ($U-B, B-V$) diagram
are shown in Fig. 3 and Fig. 4, respectively.
The $V$ magnitudes of most of the candidates are between $\sim$18 to 20 mag.
The $B-V$ colors of the candidates are between 0.3 and 1.1 and 
the $V-I$ are between 0.6 and 1.2.
Some characteristics of these GC candidates will be discussed in the next Section.

\section{DISCUSSION}

\subsection{Stellar Mass Function}

The upper part of stellar mass function (MF) of NGC 300 can be derived from the photometry of
bright stars in the galaxy. 
But as mentioned in Section III (b) the most massive part of MF
is highly uncertain because we have only the photometric data for the stars.
Due to the degeneracy of $B-V$ color for the hottest stars (T$_{eff} > $20,000K),
it is nearly impossible to estimate accurately the mass of blue stars in NGC 300.
To show this effect more clearly, we plotted two CMDs with theoretical evolution tracks
in Fig.~8. The evolution tracks in Fig.~8 (a) are those of the Padova group 
(Bertelli et al. 1994), and are transformed to the observational plane using the calibration of
Sung (1995). The model shows that most massive star (m = 100$M_\odot$) does not evolve to the
brightest part of the CMD. On the other hand, that of the Geneva group 
(Schaller et al. 1992)
shows that 120 $M_\odot$ star evolves to the brightest part of CMD. The $\eta$ Car,
the most massive star in the Galaxy
(m $\approx 100 M_\odot$) is one of the brightest stars in the Galaxy. 
This means that
at least a few of the bright blue stars in NGC 300 have masses around $100 M_\odot$.

To derive the MF of NGC 300, we should know the mass - luminosity (ML) relation. To derive
the ML relation, we used the evolution tracks by the Geneva group 
(Schaller et al. 1992) 
and the Padova group (Bertelli et al. 1994), and the calibration of Sung (1995). 
We simply divide the stars in the CMD
into 4 groups (4 color bins - I: $B-V \leq$ 0.2, II: 0.2 $< B-V \leq$ 0.8, III: 0.8 $< B-V \leq$
1.3, and IV: $B-V >$ 1.3). And we calculated the mean $M_V$ for a given model star
in each color bin, i.e. $< M_V > \equiv \int M_V (j) d\tau(j) / \int d\tau(j)$ during
the star evolves in the given color bin. The ML relation derived from the above scheme 
is strongly
dependent on the absolute magnitude which the star spends most of time for the given
color bin, i.e. for the color bin II, III, and IV the position of helium burning phase. 
The theoretical
ML relation for given color bin is presented in Fig.~9 (a). 
For the blue stars (group I),
they are actually evolved blue stars, but the theoretical ML relation
derived using the above scheme is that for main sequence (MS) stars, 
not for evolved blue (super)giants
because the star of a given mass spends most of their life time in the MS stage. 
Therefore
the theoretical ML relation for group I is inadequate for the actual case. For group
I stars, we simply determined the mid-point of evolution tracks. 
The adopted ML relation
for the group I is also presented in Fig.~9 (a).

To derive the luminosity function (LF) of each group, 
we calculated the number of stars for
each group. And we subtracted the number of GC candidates and the number
of foreground stars based on the estimates by Ratnatunga \& Bahcall (1985) for the Sculptor dwarf
galaxy. Ratnatunga \& Bahcall (1985) presented the expected number of foreground stars in the direction
of the Sculptor dwarf galaxy in $\Delta V$ = 2 mag and three bins in $B-V$ ($<$ 0.8, 0.8
-- 1.3, and $>$ 1.3 mag). We estimated the number of stars for given color
bin by keeping the trend of the variation of star number. For blue stars (group I)
most of them are bona fide members of NGC 300.
Even though there are apparently many bright stars in the group II (see Fig.~8),
the final number of stars in group II is very few. The resulting LF of stars in
NGC 300 is presented in Fig.~9 (b). 
In the transformation of the LF to the MF,
we used the ML relations derived above using the models by the Geneva group,
i.e. $\xi (\log m) d \log m \equiv N (M_V) d M_V$
(Salpeter 1955). 
The resulting MF of each group is presented in Fig.~9 (c).

To derive the initial mass function (IMF) of a heterogeneous group of stars, such
as a galaxy like NGC 300, we should know the star formation history.
It is practically impossible to know the history of star formation in NGC 300.
To derive the IMF of NGC 300, we assumed that the star formation rate is
constant with time. This assumption is valid at least for the massive stars. For this case,
the transformation of MF to IMF is very simple, i.e. only applying the difference
in the total life span between stars with different masses. By interpolating the MF
of each group, we derived the MF of bright stars in NGC 300 by summing up
the MF of each group. After that by dividing the difference in life span, we derived
the IMF of NGC 300. The total MF and IMF is presented in the same figure.
The slope of IMF ($\Gamma \equiv d \log \xi (\log m) / d \log m$) of NGC 300
is $-2.6$ $\pm$ 0.3. This value is steeper than that of the solar
neighborhood ($\Gamma = -1.35$: Salpeter 1955), but less steeper than the general
field of the magellanic clouds ($\Gamma$ = $-4.1$ $\pm$ 0.2 for LMC
and $-3.7$ $\pm$ 0.5 for SMC: Massey et al. 1995). 
The IMF for NGC 300 is that for the bright stars
measured in this study, i.e. a mixed group of stars, 
and the slope of the IMF is reasonably agreed to
the result obtained by Massey et al. (1995). 
In addition, if the star formation rate
is a decreasing function of time, then the resulting slope of IMF is steeper
than the actual value of the slope.

\subsection{The Brightest Stars}

The birghtest stars in galaxies have been used as distance indicator
(Lyo \& Lee 1997).
In order to avoid the contamination of foreground stars,
the mean brightness of three brightest stars in a galaxy is used for this
purpose. For NGC 300, the identification of red supergiant stars is very
difficult due to the possible contamination of foreground thick disk or
halo stars in the Galaxy as well as the GCs in NGC 300.
For blue stars in NGC 300, they are well-separated from the foreground
stars, and therefore they are nearly free from the contamination.

Three brightest blue stars in NGC 300 are easily selected:
 \# 50, 51, and 52 with similar magnitudes. The mean magnitude and
color of these stars are derived to  be
$<V>$ = 17.62 $\pm$ 0.01 mag, $<B-V>$ = $-0.21$ $\pm$ 0.05 mag.
Assuming the distance to NGC 300 adopted in this study,
the mean absolute magnitude of three brightest
blue stars is derived to be $<M_V^{BSG}(3)>$ = $-8.95$ mag. 
The brightness of these
blue supergiants in NGC 300 is very similar to that of three blue supergiants
in M33 (Lyo \& Lee 1997).
The calibration for the blue supergiants was given by Lyo \& Lee (1997):
$<M_V^{BSG}(3)>= 0.32 (\pm 0.05) M_B^T - 2.69 (\pm 0.94)$.
Using this calibration, the absolute magnitude of the brightest blue supergiants 
expected for the luminosity of NGC 300 ($M_B^T= -18.59$) is derived to be
$<M_V^{BSG}(3)> = -8.63 (\pm 0.54)$, which agrees within calibration
uncertainty with the value obtained in this study.

\subsection{Characteristics of the Globular Cluster Candidates}
 
\subsubsection{Luminosity function}
 
  Fig. 10 shows the GC luminosity function (GCLF) of NGC 300 obtained 
for the 17 GC candidates found in Section IV relative to the GCLF of M31 GCs.
Only class 1 objects are plotted in Fig. 10 (a),
objects of classes of 1 and 2 are plotted in Fig. 10 (b), and
objects of all three classes are plotted in Fig. 10 (c).
Milky Way Galaxy (MWG) GC data (N=146) taken from Harris (1996) are 
shown in dotted lines and
M31 GC data (N=429) taken from Barmby et al. (2000) are also shown 
in dashed lines for comparison.

  GCLF is well fitted by Gaussin or $t_5$ function
(Barmby, Huchra, \& Brodie 2001; Secker 1992; Secker \& Harris 1993).
However our GCLF of NGC 300 shown in Fig. 10 (c)
is far from Gaussin shape and clearly lacks symmetry.
Considering the fact that the faintest bin of our GCLF is 
much affected by the photometric errors and
assuming the turnover of NGC 300 GCLF is also at $M_V \sim -7.4$ 
as in the case of the Milky Way (Harris 2001),
it can be said that we have obtained the bright half of the LF.

From the largest number GCLF of NGC 300 (Fig. 10 (c))
it is concluded that the faintest bin of GCLF of NGC 300 may possibly contain
some non-GCs, and there maybe more fainter (than the turnover) GCs not 
discovered yet in NGC 300.
Further search for the rest of GCs is needed for this galaxy.

\subsubsection{Color distribution}
  We present the $B-V$ and $V-I$ color distribution of our GC candidates in Fig. 11.
The data of MWG GCs (Harris 1996; N=117 for $B-V$ and N=97 for $V-I$)
and M31 GCs (Barmby et al. 2000; N=383 for $B-V$ and N=242 for $V-I$) 
are also plotted with dotted and dashed lines, respectively.
The labels in the ordinates are for the GC candidates in NGC 300 and
are in arbitrary scales for the GCs in M31 and MWG.

In general the color distributions of the three galaxies agree well, though
there are some differences: 
(1) the peak is skewed to the blue side ($(B-V)_{peak} \sim 0.5$) in the $(B-V)$ color 
distribution (Fig. 11 (c))
and there are double peaks ($(V-I)_{peak} \sim 0.6$ and 1.2) in the $(V-I)$ color
distribution (Fig. 11 (f)), and
(2) there are more blue GC candidates in NGC 300 than red GC candidates
compared to MWG or M31.
The reason for the latter might be either ($i$) NGC 300 has actually larger rate
of blue to red clusters than MWG or M31 or ($ii$) we have discarded more 
red GC candidates than blue ones taking them to be background galaxies or 
foreground stars during the search process.
It is not easy to find out which is the cause at the present.
It is needed to perform a new, deep search for GCs in NGC 300
to find more clusters and to distinguish which is the cause.

\subsubsection{Spatial distribution}
 
  In Fig. 12 we show the histogram of the projected galactocentric distances
of the GC candidates from the nuclues of NGC 300 in arcmin.
Most candidates distribute evenly in the range 
$1' < R_{gc} < 11'$, 
while there are more objects at $4' < R_{gc} < 8'$.
  It is not clear if this underabundance of GC candidates 
in the galaxy center is real or 
due to the low discovery rate of our search in the galactic center region.


\subsection{Nucleus}

The nucleus of NGC 300 
was originally included in our GC candidates as Class 1 or 2 object,
which means it could be regarded as a star cluster in photometric images.
Based on little emission from ionized gas from the nuclear region of NGC 300,
Soffner et al. (1996) have suggested that the nucleus of NGC 300 is very 
probably an unresolved compact stellar cluster,
similar to the nucleus of M33 (Kormendy \& McClure 1993).
A detailed aperture photometry and surface photometry of the nucleus region and 
the surface photometry of the whole CCD area of NGC 300
(just like in the case of NGC 205: Kim \& Lee 1998) 
will follow in a subsequent paper(Kim et al. in preparation).

\section{SUMMARY AND CONCLUSIONS}
 
  We have presented the $UBVI$ CCD photometry of
$20.'5 \times 20.'5$ area of the Sculptor group galaxy NGC 300.
The primary results from this study of the stellar populations
and the GC candidates in this galaxy are summarized as follows.

(1) The ($V, B-V$) CMD shows well-defined blue stars 
extending up to $V \approx 17.5$ mag.

(2) The ($U-B, B-V$) diagram shows that most bright blue stars in NGC 300 
are blue supergiants.

(3) CMDs of 18 OB associations with more than 30 member stars 
are presented together with ages from the Padova isochrones.

(4) The initial mass function (IMF) of bright stars in NGC 300 
is derived to have the slope of $\Gamma = -2.6$ $\pm$ 0.3, 
which is steeper than that of the solar
neighborhood ($\Gamma = -1.35$), but less steeper than those of the general
field of the magellanic clouds ($\Gamma$ = $-4.1$ for LMC
and $-3.7$ for SMC).

(5) Assuming the distance of NGC 300 to be $(m-M)_0 = 26.53\pm 0.07$, 
the mean absolute magnitude of three brightest
blue stars is estimated to be $<M_V^{BSG}(3)>$ = $-8.95$ mag.

(6) From the search for GCs in NGC 300, we discovered 17 GC candidates
and they are 4 probable (class 1), 7 possible (class 2), and 
6 doubtful (class 3) candidates.  
Some characteristics of these newly found GC candidates are discussed.

\vspace{5mm}
This work was supported in part by the Korea Research Foundation Grant (KRF-2000-DP0450).


\begin{table*}[t]
\begin{center}
{\bf Table 1.}~~Basic Information of NGC 300 \\
\vskip 3mm
{\footnotesize
\setlength{\tabcolsep}{1.2mm}
\begin{tabular}{llc} \hline\hline
Parameter & Information & Reference \\
\hline
$\alpha_{2000}$, $\delta_{2000}$ &
        0$^h$ 54$^m$ 53.$^s$8, $-37$\arcdeg~ $40'$ $57''$ & de Vaucouleurs et al. 1991 \\
$l, b$             & 299.\arcdeg20, $-79.\arcdeg42$ & de Vaucouleurs et al. 1991 \\
Type               & SA(s)d                        & de Vaucouleurs et al. 1991 \\
{\sc{H~i}} heliocentric radial velocity, $v_\odot$ & 
				142 $\pm$ 4 km s$^{-1}$ & de Vaucouleurs et al. 1991 \\
Maximum rotational velocity, $V_m$ & 87 km s$^{-1}$ & Carignan 1985 \\
Foreground reddening, $E(B-V)$ & 0.013 mag & Schlegel et al. 1998 \\
Isophotal major diameter at the 25.0 $B$ mag arcsec$^{-2}$ level, D$_{25}$ & 
				21.$'$9 & de Vaucouleurs et al. 1991 \\
Position angle 
& 111\arcdeg  & de Vaucouleurs et al. 1991 \\
Minor to major axis ratio at D$_{25}$, $(b/a)_{25}$  &  0.74 & Carignan 1985 \\
Inclination $i$    & $42.\arcdeg3 \pm 3.\arcdeg0$ & Carignan 1985 \\
Distance modulus, (m-M)$_0$ & $26.53 \pm 0.07$ mag & Freedman et al. 2001 \\
Distance, $d$ & $2.02\pm 0.07$ Mpc (1$''$ = 9.8 pc) & Freedman et al. 2001 \\
Exponential disk corrected central surface brightness, $\mu_B(0)$ & 22.23 mag arcsec$^{-2}$ &
				Carignan 1985 \\
Scale length, $r_s$                          & 2.06 kpc            & Carignan 1985 \\
Apparent total magnitude, $B_{\rm{T}}$       & $8.72 \pm 0.05$ mag & de Vaucouleurs et al. 1991 \\
Apparent total color index, $(B-V)_{\rm{T}}$ & $0.59 \pm 0.03$ mag & de Vaucouleurs et al. 1991 \\
Apparent total color index, $(U-B)_{\rm{T}}$ & $0.11 \pm 0.03$ mag & de Vaucouleurs et al. 1991 \\
Corrected total magnitude, $B_{\rm{T}}^0$    & 8.53                & de Vaucouleurs et al. 1991 \\
Corrected total color index, $(B-V)^0_{\rm{T}}$ &  0.56            & de Vaucouleurs et al. 1991 \\
Correctedtotal color index, $(U-B)_{\rm{T}}^0$ &  0.09             & de Vaucouleurs et al. 1991 \\
Absolute total magnitude, $M_{B_{T}}$ & $-18.59$ & Sandage \& Tammann 1981 \\
Metallicity from the ratio of carbon stars to M stars of & & \\
	\hspace*{2cm} spectral type M5 or later, [Fe/H] & 
				$-0.5$ dex & Richer, Pritchet, \& Crabtree 1985 \\
{\sc{H~i}} column density in the direction of NGC 300 & $2.97 \times 10^{20}$ atoms cm$^{-2}$ &
	Read, Ponman, \& Strickland 1997 \\
{\sc{H~i}} flux            & $\leq 670$ Jy km s$^{-1}$ & Carignan 1985 \\
{\sc{H~i}} mass, $M_{H I}$ & $(2.5 \pm 0.1) \times 10^9$ $M_\odot$ & 
                                Rogstad, Crutcher, \& Chu 1979 \\
\hline
\end{tabular}
} 
\end{center}
\end{table*}
 
\begin{table*}
\begin{center}
{\small {\bf Table 2.}~Photometric data of the bright stars with $V < 19.0$ mag 
  in NGC 300$^{\rm a}$} \\
\vskip 3mm
{\scriptsize
\setlength{\tabcolsep}{1.5mm}
\begin{tabular}{rcccrrrcrrrrrrrll} \hline\hline
ID & $\alpha$ (J2000) & $\delta$ (J2000) & $V$ & $V-I$ & $B-V$ & $U-B$ &
$\epsilon_{V}$ & $\epsilon_{V-I}$ & $\epsilon_{B-V}$ & $\epsilon_{U-B}$ &
\multicolumn{4}{c}{n$_{\rm obs}$} & {V/GC$^{\rm b}$} & {remarks$^{\rm c}$} \\
\hline
  1&0:54:17.05&-37:46:50.7&13.093& 0.697& 0.744& 0.360&0.030&0.039&0.036&0.019&1&1&1&1&&\\
  2&0:55:44.10&-37:41: 5.7&13.295& 0.681& 0.635& 0.061&0.028&0.038&0.030&0.012&1&1&1&1&&\\
  3&0:55:26.36&-37:34:28.4&13.421& 0.911& 0.866& 0.613&0.023&0.028&0.024&0.008&1&1&1&1&&\\
  4&0:55: 6.44&-37:42:57.8&13.730& 0.628& 0.599& 0.076&0.010&0.013&0.011&0.009&1&1&1&1&&\\
  5&0:55:40.64&-37:40: 7.3&13.769& 0.692& 0.615& 0.078&0.011&0.021&0.013&0.011&1&1&1&1&&\\
  6&0:55:30.32&-37:33:57.5&14.175& 0.934& 0.875& 0.501&0.014&0.018&0.015&0.009&1&1&1&1&&\\
  7&0:55: 3.83&-37:51:19.7&14.216& 0.691& 0.524&-0.029&0.010&0.016&0.013&0.010&1&1&1&1&& G8 = W8\\
  8&0:54:29.49&-37:48:39.8&14.385& 1.059& 1.010& 0.864&0.008&0.013&0.009&0.016&1&1&1&1&&\\
  9&0:54:50.90&-37:51:40.7&14.441& 0.866& 0.880& 0.623&0.006&0.013&0.008&0.012&1&1&1&1&& G9 = W9\\
 10&0:55:44.43&-37:39:27.3&14.649& 0.854& 0.819& 0.442&0.009&0.020&0.011&0.013&1&1&1&1&&\\
 11&0:54:52.63&-37:49: 6.9&14.679& 0.973& 0.592& 0.181&0.008&0.016&0.010&0.010&1&1&1&1&&\\
 12&0:54:25.23&-37:35:39.5&15.065& 0.914& 0.836& 0.310&0.008&0.016&0.010&0.011&1&1&1&1&&\\
 13&0:55:45.52&-37:40:42.2&15.357& 1.039& 0.959& 0.757&0.012&0.035&0.016&0.018&1&1&1&1&&\\
 14&0:54:46.90&-37:41:40.2&15.419& 1.122& 1.094& 0.985&0.006&0.011&0.008&0.019&1&1&1&1&&\\
 15&0:54:36.53&-37:34: 9.2&15.427& 1.086& 1.051& 0.898&0.008&0.015&0.011&0.017&1&1&1&1&&\\
 16&0:55:39.86&-37:37:25.6&15.506& 0.963& 0.924& 0.722&0.012&0.024&0.014&0.016&1&1&1&1&&\\
 17&0:55:35.12&-37:41:54.6&15.597& 0.900& 0.874& 0.628&0.012&0.025&0.014&0.017&1&1&1&1&&\\
 18&0:54:49.12&-37:40:12.7&15.634& 0.679& 0.608&-0.103&0.007&0.012&0.009&0.010&1&1&1&1&&\\
 19&0:54:39.50&-37:33:20.0&15.718& 0.671& 0.591&-0.044&0.009&0.015&0.011&0.013&1&1&1&1&&\\
 20&0:55:45.62&-37:40:40.8&15.766& 1.212& 1.067& 0.895&0.013&0.026&0.016&0.024&1&1&1&1&&\\
 21&0:55:16.00&-37:37: 8.6&15.819& 0.851& 0.769& 0.274&0.007&0.013&0.009&0.013&1&1&1&1&&\\
 22&0:55:44.54&-37:41:15.1&15.884& 0.732& 0.634& 0.117&0.010&0.023&0.011&0.013&1&1&1&1&&\\
 23&0:55: 5.68&-37:44:47.1&15.966& 1.145& 1.078& 0.998&0.008&0.013&0.011&0.024&1&1&1&1&&\\
 24&0:54:39.53&-37:37:13.8&16.056& 0.814& 0.751& 0.176&0.009&0.016&0.011&0.012&1&1&1&1&&\\
 25&0:55:36.78&-37:36:23.6&16.083& 0.695& 0.625&-0.014&0.009&0.017&0.011&0.014&1&1&1&1&&\\
 26&0:54:12.60&-37:49: 3.5&16.097& 0.862& 0.681& 0.122&0.028&0.043&0.034&0.027&1&1&1&1&&\\
 27&0:55:32.53&-37:41:20.5&16.142& 0.964& 0.935& 0.715&0.008&0.015&0.010&0.020&1&1&1&1&&\\
 28&0:55:50.77&-37:50: 9.2&16.202& 1.698& 1.383& 1.282&0.013&0.021&0.021&0.048&1&1&1&1&&\\
 29&0:54:53.41&-37:41: 2.8&16.365& 0.939& 0.657& 0.094&0.047&0.074&0.059&0.047&1&1&1&1&&\\
 30&0:54:24.97&-37:44:29.9&16.369& 0.726& 0.637&-0.015&0.009&0.021&0.012&0.014&1&1&1&1&&\\
 31&0:55:19.23&-37:36:35.6&16.423& 1.283& 1.094& 0.998&0.011&0.021&0.013&0.023&1&1&1&1&&\\
 32&0:54:22.85&-37:39:31.4&16.451& 1.399& 1.247& 1.131&0.013&0.024&0.017&0.039&1&1&1&1&&\\
 33&0:54:48.43&-37:49:51.1&16.590& 0.901& 0.780& 0.321&0.011&0.025&0.014&0.019&1&1&1&1&&\\
 34&0:54:21.83&-37:32: 8.4&16.703& 0.855& 0.757& 0.333&0.019&0.031&0.022&0.021&1&1&1&1&&\\
 35&0:55: 9.32&-37:33:37.8&16.705& 1.443& 1.332& 1.171&0.008&0.019&0.013&0.060&1&1&1&1&&\\
 36&0:54:21.18&-37:38: 3.6&16.746& 0.777& 0.730& 0.047&0.014&0.024&0.017&0.019&1&1&1&1&&\\
 37&0:55:12.40&-37:45:54.6&16.795& 0.801& 0.622& 0.073&0.011&0.016&0.014&0.020&1&1&1&1&&\\
 38&0:54:26.02&-37:51:59.2&16.852& 1.037& 0.912& 0.653&0.013&0.025&0.016&0.028&1&1&1&1&&\\
 39&0:55:35.58&-37:35:41.7&16.973& 0.623& 0.423&-0.166&0.012&0.021&0.015&0.017&1&1&1&1&&\\
 40&0:55:35.13&-37:34:49.0&17.150& 0.866& 0.676& 0.038&0.015&0.030&0.018&0.023&1&1&1&1&&\\
 41&0:54:35.03&-37:46:12.4&17.196& 0.814& 0.675& 0.060&0.011&0.022&0.016&0.022&1&1&1&1&&\\
 42&0:55:26.02&-37:45:31.2&17.220& 1.169& 0.988& 0.821&0.011&0.019&0.014&0.045&1&1&1&1&&\\
 43&0:55:53.09&-37:38:27.6&17.267& 0.593& 0.415&-0.194&0.016&0.039&0.019&0.025&1&1&1&1&&\\
 44&0:54:32.37&-37:39:52.8&17.307& 0.676& 0.449&-0.225&0.012&0.022&0.015&0.018&1&1&1&1&&\\
 45&0:55: 0.96&-37:34:38.2&17.307& 2.706& 1.560& 1.294&0.012&0.025&0.019&0.097&1&1&1&1&&\\
 46&0:55:27.02&-37:32:27.6&17.327& 1.972& 1.378& 1.265&0.021&0.027&0.026&0.106&1&1&1&1&&\\
 47&0:54:39.67&-37:42:12.8&17.398& 0.697& 0.576&-0.089&0.010&0.021&0.013&0.023&1&1&1&1&&\\
 48&0:54:37.50&-37:46: 2.6&17.504& 1.615& 1.377& 1.350&0.010&0.021&0.020&0.093&1&1&1&1&&\\
 49&0:54:53.72&-37:32:24.3&17.509& 2.091& 1.436& 0.972&0.013&0.019&0.018&0.056&1&1&1&1&&\\
 50&0:55: 3.65&-37:43:19.2&17.618&-0.272&-0.226&-1.166&0.048&0.082&0.065&0.064&1&1&1&1&&\\
 51&0:55:13.54&-37:41:37.4&17.624&-0.026&-0.143&-0.916&0.021&0.035&0.028&0.055&1&1&1&1&&\\
 52&0:55:12.22&-37:41:19.2&17.625&\nodata&-0.232&-0.850&0.042&\nodata&0.053&0.059&1&0&1&1&&\\
 53&0:55:54.93&-37:43:32.3&17.627& 1.292& 1.144& 1.041&0.018&0.032&0.027&0.077&1&1&1&1&&\\
 54&0:54:30.27&-37:37:55.5&17.668& 0.165& 0.089&-0.556&0.012&0.026&0.015&0.015&1&1&1&1&&\\
 55&0:55:30.72&-37:45:43.0&17.674& 1.200& 1.031& 0.846&0.013&0.024&0.017&0.051&1&1&1&1&&\\
 56&0:54:52.60&-37:41:48.2&17.678& 1.898& 1.426& 1.216&0.012&0.021&0.018&0.129&1&1&1&1&&\\
 57&0:54:50.99&-37:52: 2.9&17.689& 1.307& 1.123& 1.052&0.012&0.019&0.018&0.065&1&1&1&1&& W29\\
 58&0:55: 1.00&-37:34:39.9&17.702& 2.900& 1.575& 1.026&0.013&0.021&0.025&0.115&1&1&1&1&&\\
 59&0:54:42.79&-37:43: 0.9&17.856&-0.215&-0.020&-1.026&0.041&0.079&0.048&0.045&1&1&1&1&&\\
 60&0:54:35.80&-37:43: 5.6&17.878& 0.769& 0.561&-0.156&0.034&0.046&0.056&0.055&1&1&1&1&g2&\\
 61&0:54:13.22&-37:34:19.3&17.880& 0.686& 0.555& 0.100&0.023&0.037&0.030&0.090&1&1&1&1&&\\
 62&0:55:13.74&-37:41:39.5&17.886& 0.390& 0.317&-0.333&0.023&0.040&0.044&0.053&1&1&1&1&&\\
 63&0:55:20.28&-37:48: 7.1&17.927& 3.081& 1.684& 1.301&0.013&0.018&0.025&0.157&1&1&1&1&&\\
 64&0:55:51.13&-37:37: 3.6&17.954& 1.556& 1.296& 0.815&0.017&0.031&0.025&0.111&1&1&1&1&&\\
 65&0:54:42.77&-37:40: 3.1&17.984& 0.197& 0.011&-1.084&0.022&0.036&0.030&0.033&1&1&1&1&&\\
 66&0:54:22.97&-37:34:51.3&18.016& 0.120& 0.051&-0.496&0.024&0.042&0.031&0.028&1&1&1&1&&\\
 67&0:55:13.77&-37:37:49.1&18.025& 0.362& 0.065&-0.943&0.016&0.032&0.018&0.015&1&1&1&1&&\\
 68&0:54:40.32&-37:40:53.9&18.027& 0.063&-0.118&-1.080&0.040&0.059&0.049&0.052&1&1&1&1&&\\
 69&0:54:20.27&-37:34: 2.6&18.060& 1.022& 0.968& 0.668&0.043&0.195&0.051&0.191&1&1&1&1&&\\
 70&0:54:21.92&-37:38:14.0&18.091& 0.139& 0.084&-0.693&0.013&0.041&0.019&0.022&1&1&1&1&&\\
 71&0:55:49.97&-37:33:18.3&18.103& 0.915& 0.589& 0.049&0.055&0.080&0.082&0.073&1&1&1&1&&\\
 72&0:55:11.03&-37:36:46.1&18.158& 1.308& 0.601& 0.321&0.108&0.122&0.119&0.074&1&1&1&1&g1&\\
 73&0:55:12.91&-37:41:37.9&18.192&-0.135&-0.140&-1.011&0.043&0.061&0.059&0.057&1&1&1&1&&\\
\hline
\end{tabular}
} 
\end{center}
\end{table*}
\begin{table*}
\begin{center}
{\small {\bf Table 2.}~-~Continued }
\vskip 3mm
{\scriptsize
\setlength{\tabcolsep}{1.5mm}
\begin{tabular}{rcccrrrcrrrrrrrll} \hline\hline
ID & $\alpha$ (J2000) & $\delta$ (J2000) & $V$ & $V-I$ & $B-V$ & $U-B$ &
$\epsilon_{V}$ & $\epsilon_{V-I}$ & $\epsilon_{B-V}$ & $\epsilon_{U-B}$ &
\multicolumn{4}{c}{n$_{\rm obs}$} & {V/GC$^{\rm b}$} & {remarks$^{\rm c}$} \\
\hline
 74&0:55: 2.56&-37:38:29.1&18.193& 0.213& 0.126&-0.192&0.015&0.044&0.022&0.036&1&1&1&1&&\\
 75&0:55:35.60&-37:39:34.6&18.193& 0.688& 0.483&-0.210&0.015&0.035&0.019&0.037&1&1&1&1&&\\
 76&0:54:51.57&-37:39:21.0&18.194& 1.266& 1.018& 0.704&0.015&0.023&0.025&0.078&1&1&1&1&&\\
 77&0:54:53.10&-37:32:49.5&18.208& 1.056& 0.945& 0.756&0.011&0.022&0.019&0.079&1&1&1&1&&\\
 78&0:54:49.85&-37:40:10.5&18.224& 0.720& 0.665& 0.222&0.018&0.031&0.028&0.061&1&1&1&1&&\\
 79&0:55: 0.83&-37:46:59.0&18.290& 0.866& 0.686& 0.147&0.015&0.033&0.021&0.053&1&1&1&1&&\\
 80&0:54:39.40&-37:38:47.5&18.301& 0.296& 0.113&-0.482&0.013&0.032&0.021&0.024&1&1&1&1&&\\
 81&0:54:25.42&-37:39: 8.5&18.302& 0.301&-0.029&-0.804&0.032&0.043&0.040&0.039&1&1&1&1&&\\
 82&0:54:12.69&-37:52:19.0&18.324& 0.788& 0.723& 0.123&0.031&0.052&0.040&0.057&1&1&1&1&&\\
 83&0:55:12.61&-37:41:39.2&18.328&-0.270&-0.076&-1.234&0.079&0.136&0.132&0.135&1&1&1&1&&\\
 84&0:55:24.96&-37:50:56.1&18.360& 0.718& 0.590& 0.042&0.016&0.030&0.021&0.051&1&1&1&1&&\\
 85&0:54:30.89&-37:49:26.3&18.390& 0.793& 0.603&-0.079&0.013&0.034&0.023&0.036&1&1&1&1&&\\
 86&0:55:12.68&-37:41:39.4&18.393&-0.106&-0.213&-1.081&0.085&0.139&0.133&0.118&1&1&1&1&&\\
 87&0:54:26.68&-37:44: 3.5&18.396& 0.087& 0.097&-0.453&0.016&0.035&0.021&0.028&1&1&1&1&&\\
 88&0:55:38.35&-37:41:57.7&18.398& 0.238& 0.098&-0.719&0.020&0.028&0.024&0.022&1&1&1&1&&\\
 89&0:54:50.26&-37:38:23.1&18.405&-0.262&-0.140&-0.939&0.040&0.082&0.052&0.060&1&1&1&1&&\\
 90&0:55:39.42&-37:43:22.3&18.409& 0.817& 0.581&-0.035&0.016&0.037&0.034&0.045&1&1&1&1&&\\
 91&0:54:17.18&-37:48: 7.5&18.416& 0.641& 0.432&-0.197&0.019&0.032&0.024&0.030&1&1&1&1&&\\
 92&0:54:58.25&-37:41:13.1&18.418& 0.890& 0.702& 0.170&0.016&0.026&0.021&0.046&1&1&1&1&&\\
 93&0:55:12.45&-37:41:19.5&18.433&-0.010&-0.071&-0.889&0.021&0.041&0.025&0.027&1&1&1&1&&\\
 94&0:54:25.48&-37:39:36.3&18.437&-0.070&-0.052&-0.572&0.028&0.047&0.038&0.049&1&1&1&1&&\\
 95&0:54:30.01&-37:50: 5.0&18.441& 1.366& 1.596& 0.495&0.074&0.108&0.105&0.147&1&1&1&1&&\\
 96&0:54:14.54&-37:31:53.3&18.451& 1.067& 0.930& 0.585&0.040&0.053&0.080&0.120&1&1&1&1&&\\
 97&0:54:42.60&-37:40: 2.9&18.469& 0.011&-0.089&-0.981&0.030&0.057&0.041&0.038&1&1&1&1&&\\
 98&0:55:29.50&-37:36:43.1&18.492& 2.207& 1.523&\nodata&0.012&0.022&0.021&\nodata&1&1&1&0&&\\
 99&0:55: 2.38&-37:38:26.8&18.497&-0.157&-0.178&-1.040&0.032&0.059&0.047&0.051&1&1&1&1&&\\
100&0:54:42.69&-37:42:57.8&18.518& 1.395& 0.308&-0.920&0.055&0.061&0.063&0.055&1&1&1&1&&\\
101&0:54:49.92&-37:40:24.7&18.523& 0.575& 0.480& 0.103&0.021&0.036&0.026&0.052&1&1&1&1&&\\
102&0:54:43.37&-37:36:26.5&18.530& 0.100& 0.090&-0.579&0.015&0.035&0.021&0.024&1&1&1&1&&\\
103&0:54:54.61&-37:45:46.7&18.552& 0.201& 0.019&-0.809&0.015&0.036&0.019&0.023&1&1&1&1&&\\
104&0:55:13.16&-37:41:43.2&18.554& 0.360& 0.088&-0.891&0.024&0.047&0.034&0.041&1&1&1&1&&\\
105&0:54:32.07&-37:37:49.7&18.564& 0.186& 0.044&-0.813&0.015&0.039&0.019&0.024&1&1&1&1&&\\
106&0:55:14.73&-37:38: 7.0&18.565& 1.517& 1.259& 1.317&0.015&0.022&0.029&0.148&1&1&1&1&&\\
107&0:54:25.51&-37:39:55.1&18.586& 1.399& 1.242&-0.276&0.029&0.035&0.070&0.095&1&1&1&1&&\\
108&0:55: 3.73&-37:42:51.2&18.606&-0.105&-0.146&-1.019&0.028&0.056&0.037&0.042&1&1&1&1&&\\
109&0:55: 0.90&-37:39:30.2&18.615& 0.538& 0.347& 0.298&0.022&0.045&0.027&0.056&1&1&1&1&&\\
110&0:55:12.07&-37:41:25.2&18.615&-0.268&-0.096&-0.818&0.037&0.080&0.048&0.043&1&1&1&1&&\\
111&0:54:32.47&-37:43:37.2&18.616& 0.275& 0.173&-0.422&0.017&0.045&0.034&0.040&1&1&1&1&&\\
112&0:54:48.71&-37:44: 3.2&18.624& 0.907& 0.746& 0.209&0.018&0.035&0.026&0.075&1&1&1&1&&\\
113&0:55:13.97&-37:41:42.6&18.629& 0.667&-0.193&-0.915&0.034&0.047&0.043&0.041&1&1&1&1&&\\
114&0:54:56.90&-37:41:33.3&18.637&-0.115&-0.081&-0.895&0.022&0.056&0.028&0.025&1&1&1&1&&\\
115&0:54:16.03&-37:35:36.2&18.641& 0.965& 0.745& 0.235&0.046&0.064&0.060&0.078&1&1&1&1&&\\
116&0:54:49.68&-37:38:37.1&18.646& 0.426&-0.126&-0.953&0.026&0.048&0.040&0.043&1&1&1&1&&\\
117&0:54:52.99&-37:41:57.9&18.647& 0.770& 0.505&-0.221&0.019&0.036&0.024&0.037&1&1&1&1&&\\
118&0:54:47.07&-37:40:14.6&18.653& 0.901& 0.456&-0.453&0.025&0.041&0.033&0.047&1&1&1&1&&\\
119&0:54:59.72&-37:36:49.8&18.654& 0.851& 0.721&-0.120&0.021&0.033&0.027&0.062&1&1&1&1&&\\
120&0:55: 5.82&-37:41: 5.8&18.692& 0.672& 0.716& 0.301&0.021&0.034&0.030&0.083&1&1&1&1&&\\
121&0:54:57.41&-37:33:55.3&18.705& 1.173& 1.116& 0.348&0.077&0.101&0.101&0.113&1&1&1&1&g2&\\
122&0:54:44.10&-37:51:54.1&18.710& 1.249& 1.315& 0.406&0.067&0.103&0.099&0.136&1&1&1&1&&\\
123&0:54:55.99&-37:41:11.6&18.712& 0.031&-0.102&-0.816&0.022&0.059&0.028&0.029&1&1&1&1&&\\
124&0:54:20.83&-37:38:48.9&18.712& 0.617& 0.471& 0.147&0.020&0.036&0.028&0.050&1&1&1&1&&\\
125&0:55:53.06&-37:45:13.3&18.713& 1.136& 0.963& 0.565&0.017&0.033&0.029&0.086&1&1&1&1&&\\
126&0:55:42.39&-37:34:59.1&18.715& 1.747& 1.274& 0.848&0.015&0.033&0.028&0.200&1&1&1&1&&\\
127&0:54:40.16&-37:39:25.9&18.718& 0.238& 0.027&-0.485&0.018&0.045&0.022&0.030&1&1&1&1&&\\
128&0:54:20.29&-37:34: 1.9&18.725& 0.800& 2.333&-0.846&0.108&0.180&0.198&0.383&1&1&1&1&&\\
129&0:54:18.34&-37:35: 2.8&18.727& 1.636& 1.326&-0.365&0.020&0.034&0.039&0.086&1&1&1&1&&\\
130&0:54:22.44&-37:43:12.4&18.728& 1.968& 1.383&\nodata&0.019&0.025&0.031&\nodata&1&1&1&0&&\\
131&0:55:32.41&-37:34:18.7&18.736&-0.281&-0.247&-0.994&0.021&0.061&0.024&0.022&1&1&1&1&&\\
132&0:54:42.84&-37:42:59.7&18.743&-0.055&-0.255& 0.110&0.055&0.110&0.081&0.145&1&1&1&1&&\\
133&0:54:45.62&-37:40:16.7&18.751& 0.062& 0.061&-0.570&0.022&0.060&0.027&0.031&1&1&1&1&&\\
134&0:55:12.23&-37:41:20.9&18.752&-0.054&-0.141&-1.058&0.062&0.098&0.086&0.134&1&1&1&1&&\\
135&0:55:12.24&-37:41:20.3&18.754& 0.895& 0.070&-0.942&0.104&0.131&0.148&0.112&1&1&1&1&&\\
136&0:55:20.80&-37:48:11.2&18.755& 3.187& 1.642&\nodata&0.018&0.024&0.035&\nodata&1&1&1&0&&\\
137&0:54:53.71&-37:41: 1.1&18.756& 0.072& 0.050&-0.532&0.030&0.081&0.036&0.033&1&1&1&1&&\\
138&0:55: 3.83&-37:43:18.2&18.764&-0.195&-0.185&-0.894&0.080&0.142&0.102&0.093&1&1&1&1&&\\
139&0:55: 4.75&-37:34: 9.2&18.769& 1.600& 1.424& 0.771&0.014&0.022&0.030&0.176&1&1&1&1&&\\
140&0:54:53.35&-37:39: 1.2&18.800& 0.337& 0.125&-0.516&0.022&0.048&0.026&0.034&1&1&1&1&&\\
141&0:55:22.72&-37:33:40.1&18.803& 0.020&-0.094&-0.903&0.023&0.061&0.026&0.026&1&1&1&1&&\\
142&0:54:55.86&-37:40:14.3&18.808&-0.148&-0.195&-0.952&0.037&0.063&0.045&0.036&1&1&1&1&&\\
143&0:54:34.66&-37:37:22.9&18.810& 0.382& 0.109&-0.225&0.063&0.112&0.082&0.081&1&1&1&1&&\\
144&0:54:28.23&-37:37:54.5&18.833& 1.504& 1.725&\nodata&0.017&0.028&0.039&\nodata&1&1&1&0&&\\
145&0:55:24.77&-37:39:34.4&18.835&-0.260&-0.195&-1.103&0.051&0.084&0.070&0.061&1&1&1&1&&\\
146&0:55:13.13&-37:41:39.4&18.836&-0.999&-0.240&-0.998&0.064&0.229&0.085&0.074&1&1&1&1&&\\
\hline
\end{tabular}
} 
\end{center}
\end{table*}
\begin{table*}
\begin{center}
{\small {\bf Table 2.}~-~Continued }
\vskip 3mm
{\scriptsize
\setlength{\tabcolsep}{1.5mm}
\begin{tabular}{rcccrrrcrrrrrrrll} \hline\hline
ID & $\alpha$ (J2000) & $\delta$ (J2000) & $V$ & $V-I$ & $B-V$ & $U-B$ &
$\epsilon_{V}$ & $\epsilon_{V-I}$ & $\epsilon_{B-V}$ & $\epsilon_{U-B}$ &
\multicolumn{4}{c}{n$_{\rm obs}$} & {V/GC$^{\rm b}$} & {remarks$^{\rm c}$} \\
\hline
147&0:54:45.62&-37:43:41.0&18.839& 0.973& 0.755& 0.197&0.044&0.064&0.059&0.084&1&1&1&1&g2&\\
148&0:54:50.40&-37:40: 3.6&18.842& 0.082& 0.074&-0.647&0.039&0.186&0.072&0.094&1&1&1&1&&\\
149&0:55:27.93&-37:44:19.4&18.844& 0.667& 0.191&-0.977&0.017&0.038&0.024&0.025&1&1&1&1&&\\
150&0:54:43.44&-37:45:32.6&18.861& 0.487& 0.322&-0.055&0.066&0.078&0.071&0.107&1&1&1&1&&\\
151&0:54:43.43&-37:43: 8.8&18.872& 0.037& 0.511&-2.105&0.079&0.127&0.151&0.165&1&1&1&1&&\\
152&0:55:24.24&-37:37:28.3&18.873& 1.012& 0.848& 0.226&0.020&0.039&0.030&0.078&1&1&1&1&&\\
153&0:55:38.79&-37:46:30.7&18.884& 1.206& 1.074& 0.267&0.091&0.137&0.122&0.133&1&1&1&1&g3&\\
154&0:54:57.44&-37:44: 6.4&18.894& 0.100&-0.036&-0.693&0.024&0.049&0.036&0.036&1&1&1&1&&\\
155&0:54:40.03&-37:45:46.5&18.894& 1.153& 0.978& 1.152&0.018&0.035&0.032&0.167&1&1&1&1&&\\
156&0:55:12.76&-37:41:43.7&18.896&-0.234&-0.100&-1.017&0.078&0.124&0.105&0.121&1&1&1&1&&\\
157&0:55:27.56&-37:46: 9.4&18.905& 1.213& 1.065& 0.723&0.085&0.136&0.123&0.178&1&1&1&1&g3&\\
158&0:54:26.13&-37:43:56.9&18.921& 0.132& 0.090&-0.500&0.020&0.053&0.025&0.023&1&1&1&1&&\\
159&0:55: 3.87&-37:42:50.5&18.944&-0.150&-0.202&-0.932&0.033&0.064&0.044&0.046&1&1&1&1&&\\
160&0:54:44.75&-37:42:39.0&18.950& 0.124&-0.059&-1.034&0.017&0.057&0.024&0.025&1&1&1&1&&\\
161&0:55:17.61&-37:38:33.3&18.953& 1.585& 1.316&\nodata&0.021&0.032&0.034&\nodata&1&1&1&0&&\\
162&0:55:12.69&-37:41:38.2&18.963&-0.195&-0.375&-0.811&0.099&0.159&0.162&0.154&1&1&1&1&&\\
163&0:55:30.54&-37:40:27.9&18.965& 1.328& 1.122& 0.581&0.094&0.140&0.127&0.133&1&1&1&1&&\\
164&0:54:41.79&-37:38:35.8&18.977& 0.249& 0.008&-0.956&0.033&0.148&0.039&0.040&1&1&1&1&&\\
165&0:55:36.73&-37:44: 3.9&18.983& 2.553& 1.413&\nodata&0.021&0.028&0.044&\nodata&1&1&1&0&&\\
166&0:55:34.45&-37:41:24.7&18.986& 1.238& 1.028& 0.820&0.023&0.038&0.033&0.174&1&1&1&1&&\\
167&0:55:31.98&-37:41:39.1&18.992&-0.059& 0.107&-0.341&0.020&0.065&0.031&0.038&1&1&1&1&&\\
\hline
\end{tabular}
} 
\end{center}
\hskip 1.2cm
{\begin{minipage}{16.0cm}
$^{\rm a}$ Units of right ascension
are hours, minutes, and seconds, and units of declination are degrees,
arcminutes, and arcseconds. \\
$^{\rm b}$ V: variable star from Graham (1984), GC(g\#): GC
candidate with its class (\#: 1--3 from higher probability to lower probability) \\
$^{\rm c}$ G: Graham (1981); W: Walker (1995)
\end{minipage}  }
\end{table*}

\begin{table*}
\begin{center}
{\bf Table 3.}~~Mean photometric errors \\
\vskip 3mm
\begin{tabular}{rcrcrcrcr} \hline\hline
{$V$} & {$\epsilon_{V}$} & N & {$\epsilon_{V-I}$} &
N & {$\epsilon_{B-V}$} & N & {$\epsilon_{U-B}$} & N \\
\hline
 13.5 &  0.020 &    5 &  0.028 &    5 &  0.023 &    5 &  0.012 &    5 \\
 14.5 &  0.009 &    6 &  0.016 &    6 &  0.011 &    6 &  0.012 &    6 \\
 15.5 &  0.009 &   12 &  0.019 &   12 &  0.012 &   12 &  0.016 &   12 \\
 16.5 &  0.015 &   16 &  0.026 &   16 &  0.019 &   16 &  0.027 &   16 \\
 17.5 &  0.019 &   26 &  0.032 &   25 &  0.027 &   26 &  0.065 &   26 \\
 18.5 &  0.035 &  102 &  0.065 &  102 &  0.051 &  102 &  0.074 &   96 \\
 19.5 &  0.050 &  370 &  0.107 &  357 &  0.074 &  369 &  0.090 &  312 \\
 20.5 &  0.076 & 1381 &  0.178 & 1190 &  0.112 & 1367 &  0.123 &  966 \\
 21.5 &  0.128 & 4175 &  0.238 & 2304 &  0.182 & 3880 &  0.152 &  872 \\
\hline
\end{tabular}
\end{center}
\end{table*}

\begin{table*}[t]
\begin{center}
{\bf Table 4.}~~GC candidates \\
\vskip 3mm
{\footnotesize 
\setlength{\tabcolsep}{1.5mm}
\begin{tabular}{rcrcccrccccrrrrc} \hline\hline
ID & $\alpha$ (J2000) & $\delta$ (J2000) & $V$ & $V-I$ & $B-V$ & $U-B$ &
$\epsilon_{V}$ & $\epsilon_{V-I}$ & $\epsilon_{B-V}$ & $\epsilon_{U-B}$ &
\multicolumn{4}{c}{n$_{\rm obs}$} & class \\
\hline
1&0:54:22.07&-37:43:22.1&19.634& 1.288& 1.231& 0.192& 0.066&0.091&0.096&0.171& 1&1&1&1&1 \\
2&0:54:35.80&-37:43:05.6&17.878& 0.769& 0.561&-0.156& 0.034&0.046&0.056&0.055& 1&1&1&1&2 \\
3&0:54:40.02&-37:49:18.5&19.571& 0.868& 0.720&-0.092& 0.086&0.142&0.117&0.117& 1&1&1&1&3 \\
4&0:54:45.62&-37:43:41.0&18.839& 0.973& 0.755& 0.197& 0.044&0.064&0.059&0.084& 1&1&1&1&2 \\
5&0:54:49.35&-37:43:30.5&19.076& 0.786& 0.621& 0.191& 0.056&0.085&0.071&0.093& 1&1&1&1&2 \\
6&0:54:51.67&-37:39:38.0&19.073&\nodata& 0.444&-2.349& 0.167&\nodata&0.228&0.178& 1&0&1&1&3 \\
7&0:54:53.22&-37:43:11.1&19.213& 1.639& 0.765&-0.080& 0.164&0.176&0.192&0.130& 1&1&1&1&1 \\
8&0:54:57.41&-37:33:55.3&18.705& 1.173& 1.116& 0.348& 0.077&0.101&0.101&0.113& 1&1&1&1&2 \\
9&0:54:57.57&-37:36:37.0&19.728& 1.552& 0.665&-0.409& 0.043&0.053&0.064&0.132& 1&1&1&1&2 \\
10&0:54:59.03&-37:36:24.4&19.188& 0.557& 0.380&-0.244& 0.048&0.068&0.056&0.055& 1&1&1&1&3 \\
11&0:55:01.44&-37:37:21.8&19.739& 0.955& 0.714& 0.172& 0.078&0.118&0.096&0.152& 1&1&1&1&1 \\
12&0:55:11.03&-37:36:46.1&18.158& 1.308& 0.601& 0.321& 0.108&0.122&0.119&0.074& 1&1&1&1&1 \\
13&0:55:11.17&-37:33:13.9&19.537& 0.724& 0.609&-0.151& 0.090&0.141&0.122&0.115& 1&1&1&1&3 \\
14&0:55:17.67&-37:36:57.9&19.831& 1.166& 1.094& 0.174& 0.091&0.137&0.144&0.206& 1&1&1&1&2 \\
15&0:55:21.81&-37:45:40.1&19.625& 0.608& 0.423& 0.089& 0.073&0.125&0.100&0.117& 1&1&1&1&2 \\
16&0:55:27.56&-37:46:09.4&18.905& 1.213& 1.065& 0.723& 0.085&0.136&0.123&0.178& 1&1&1&1&3 \\
17&0:55:38.79&-37:46:30.7&18.884& 1.206& 1.074& 0.267& 0.091&0.137&0.122&0.133& 1&1&1&1&3 \\
\hline
\end{tabular}
} 
\end{center}
\end{table*}

\begin{figure*}[1]
\plotone{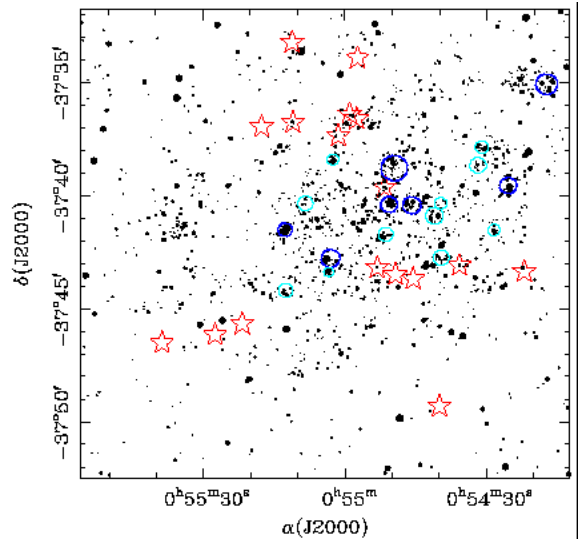}
\figcaption{Finding chart for the stars measured from
this observation ($20.'5 \times 20.'5$). To show the distribution of 
bright stars in NGC 300,
we plotted only stars with $V$ = 16.5 -- 21 mag.
The size of the dot is proportional to the magnitude of the star.
Open circles represent rich OB associations (N$_* \geq$ 30) in
NGC 300 listed in Pietrzy\'nski et al. (2001). 
Thin circles are OB associations with 30 $\leq$ N$_* <$ 50 and
thick circles are those with N$\geq$ 50.
The star symbols indicate the positions of GC candidates 
(see Section IV).
}
\end{figure*}

\begin{figure*}[2]
\plotone{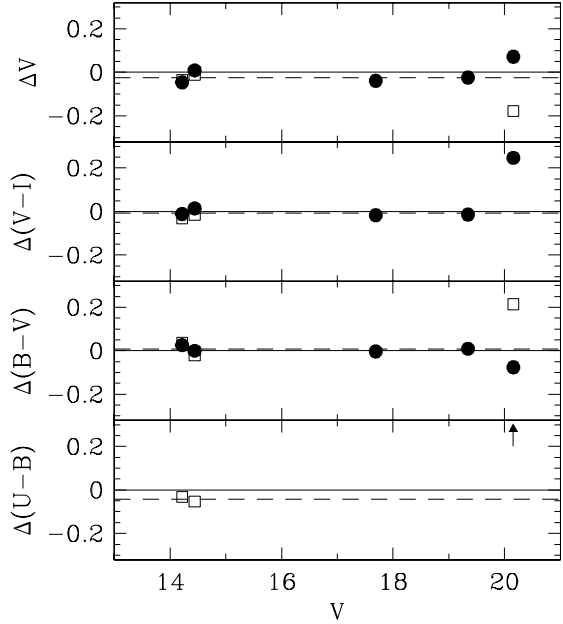}
\figcaption{Comparison of our photometry with that of 
Graham (1981)
(filled circle) and with that of Walker (1995) (open square), showing
differences in the sense, previous photometry minus our new photometry.
The arrow symbol in the $\Delta (U-B)$ plot denotes a value well outside 
of the plot.
}
\end{figure*}

\begin{figure*}[3]
\plotone{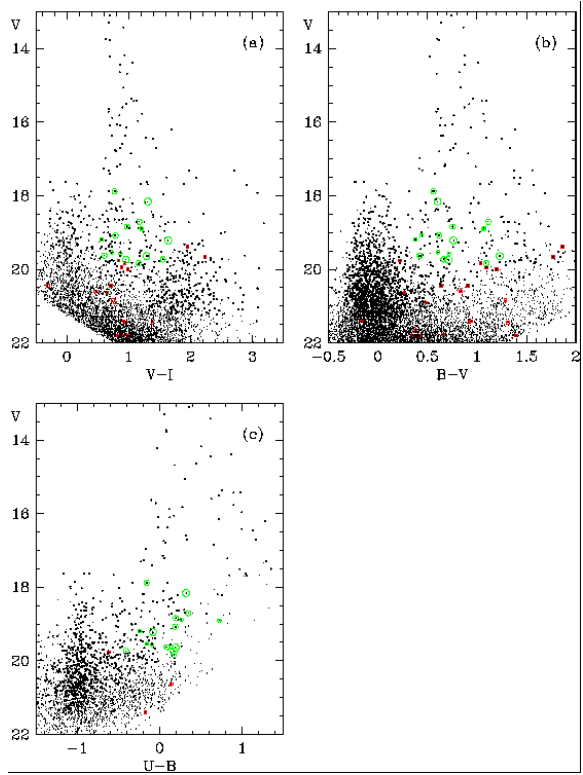}
\figcaption{Color-magnitude diagrams.
Large and small dots represent, respectively, stars with 
good photometry ($\epsilon \leq 0.1$) and bad quality ($\epsilon > 0.1$). 
The open squares with dots
and open circles with dots denote the known variable stars (Graham 1984) and 
the GC candidates, respectively. 
The size of the circle is propotional to the probability
of GC in the sense that larger for higher probability, smaller for lower.
}
\end{figure*}

\begin{figure*}[4]
\plotone{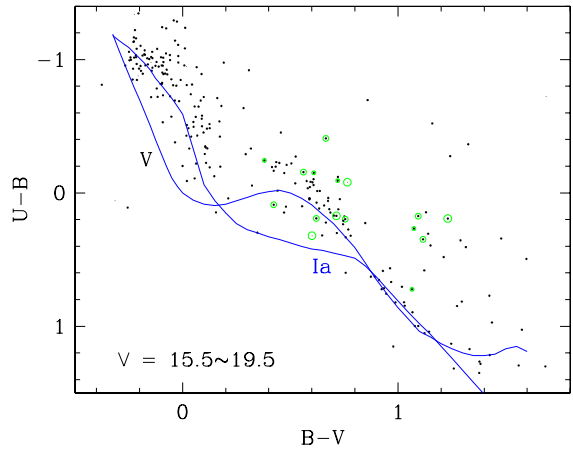}
\figcaption{$U-B$ versus $B-V$ diagram for 
$V$ = 15.5 -- 19.5 mag.
The thin and thick solid lines represent the intrinsic 
zero-age main sequence (ZAMS)
relation and the intrinsic relation for supergiants (Ia), respectively.
Large and small dots represent, respectively, stars with 
good photometry ($\epsilon \leq 0.1$) and bad quality ($\epsilon > 0.1$). 
The open circles with dots at center denote the 
GC candidates. The size of the circle is propotional to the probability
of GC in the sense that larger for higher probability, smaller for lower.
}
\end{figure*}

\begin{figure*}[5]
\plotone{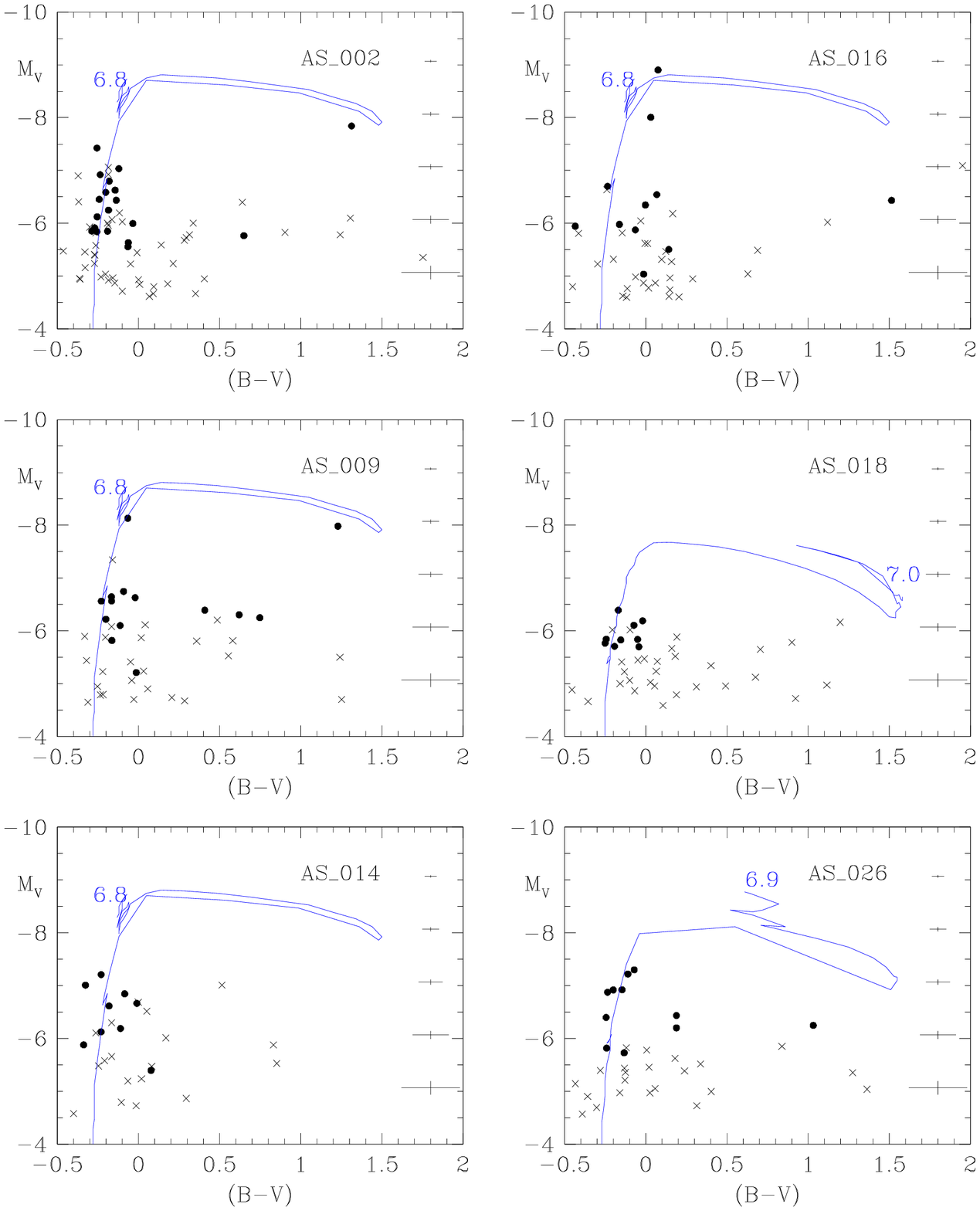}
\plotone{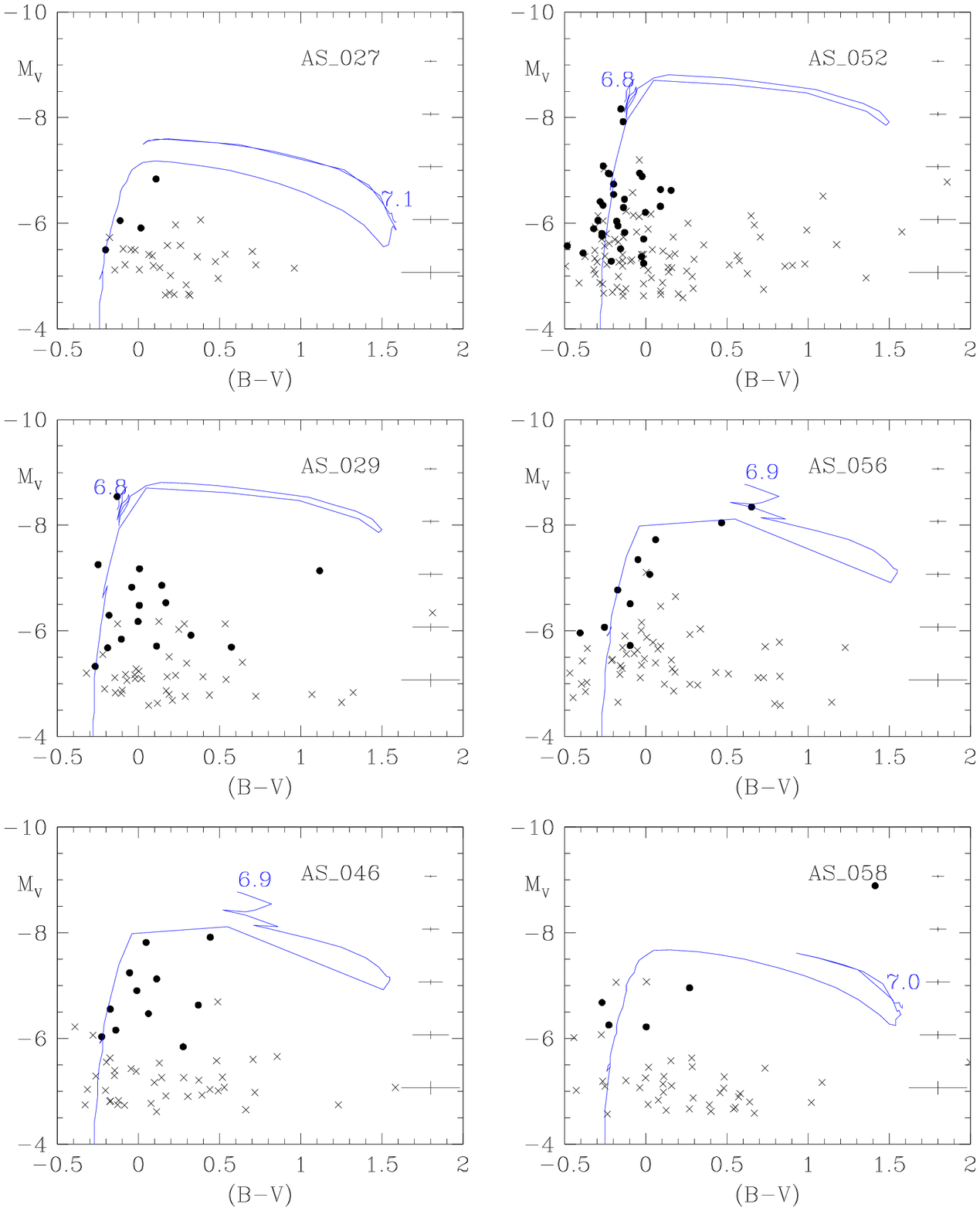}
\plotone{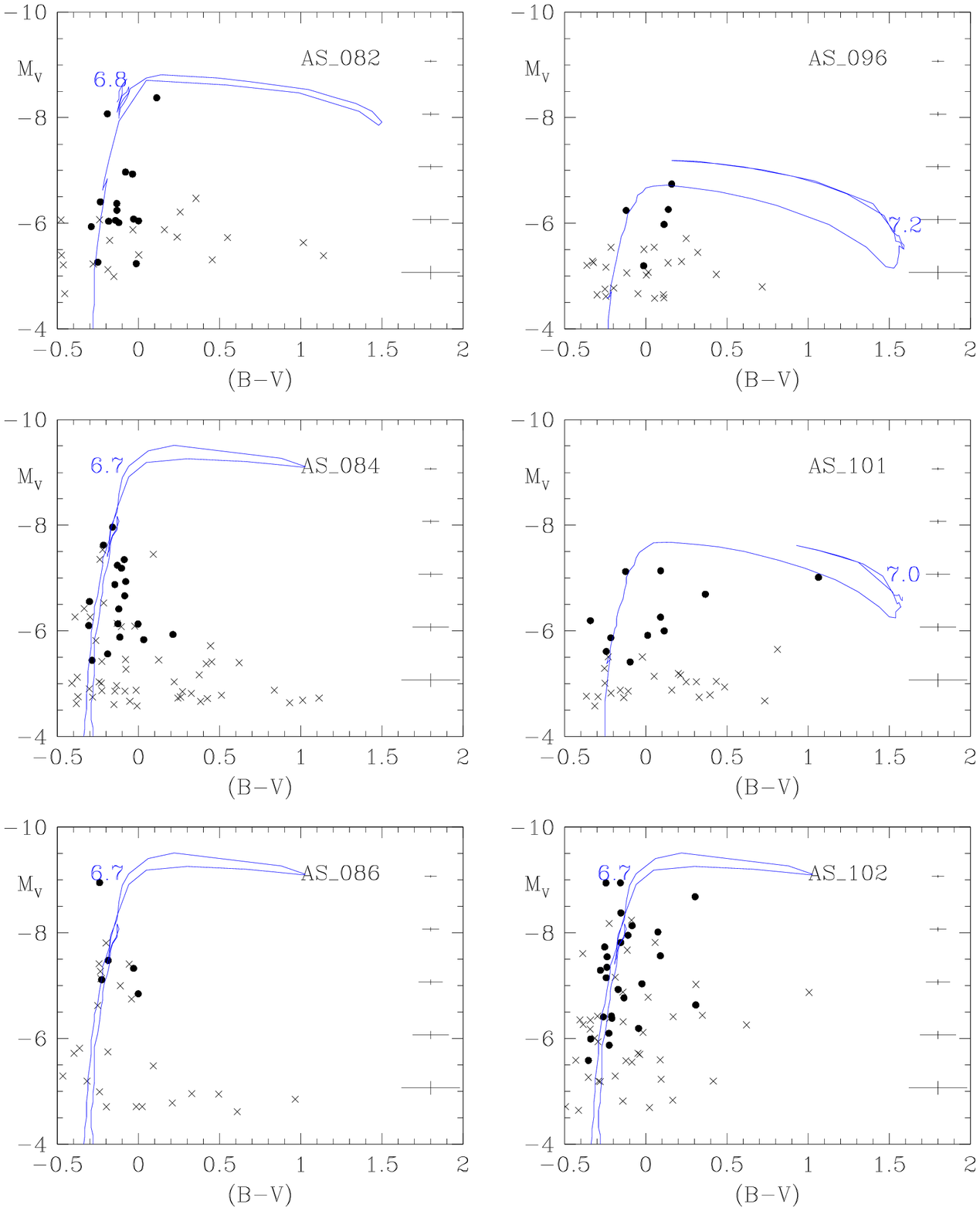}
\figcaption{Color-magnitude diagrams of selected OB associations.
The designation of OB associations is from Pietrzy\'nski et al. (2001). The dot and cross
denote, respectively, a star with good photometry ($\epsilon \leq 0.1$) and 
bad quality ($\epsilon > 0.1$). 
The mean photometric error in $V$ and $B-V$ is marked 
in the right part of each diagram.
The solid lines represent the isochrones from Bertelli et al. (1994) with the age
of the association in a logarithmic scale ($\log \tau_{age}$).
}
\end{figure*}

\begin{figure*}[6]
\plotone{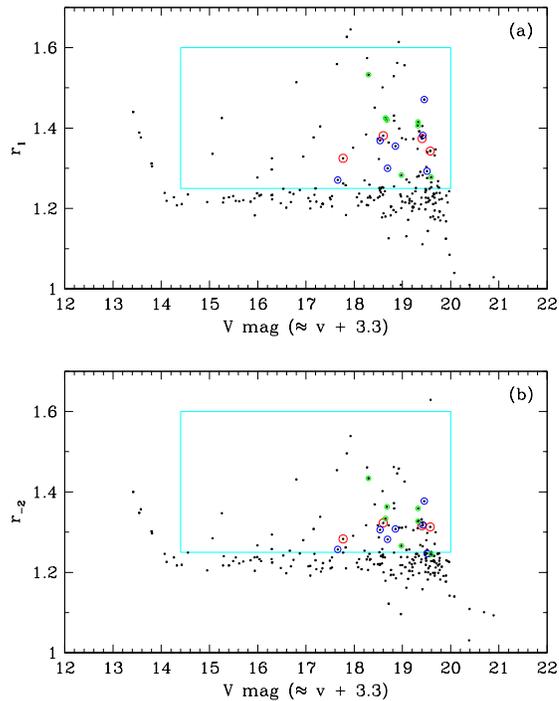}
\figcaption{The distribution of 
morphological image classification parameters 
((a) $r_1$ and (b) $r_{-2}$)
calculated on the $V$-band image are 
plotted over the $V$ magnitude.
All objects with $V<20$ and $0.3<B-V<2.0$ are plotted (N=223).
The horizontal sequence of objects are mainly stars,
while GCs in NGC 300 have a little larger moment values.
Some objects with largest moment values are probably background galaxies.
The objects in the box region are visually examined on the image
and the circled objects denote the resulting GC candidates.
The larger the circle, the higher the probability to be a GC.
}
\end{figure*}

\begin{figure*}[7]
\plotone{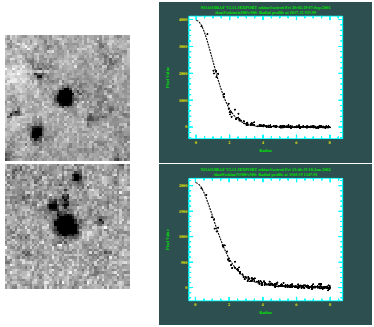}
\figcaption{$V$-band mosaic images
($30.''1 \times 30.''1$)
and radial profiles of a typical star (upper panels, Table 2 ID=47,
$V$=17.398, $B-V$=0.576)
and a class 1 GC candidate (lower panels, GC candidate ID=12, Table 2 ID=72,
$V$=18.158, $B-V$=0.601).
North is to the down, and east is to the right on the images.
The typical FWHM of stars in the $V$ image is $\sim 2.4$ pixel,
while the FWHM of this GC candidate is $\sim$ 2.8 pixel.
}
\end{figure*}

\begin{figure*}[8]
\plotone{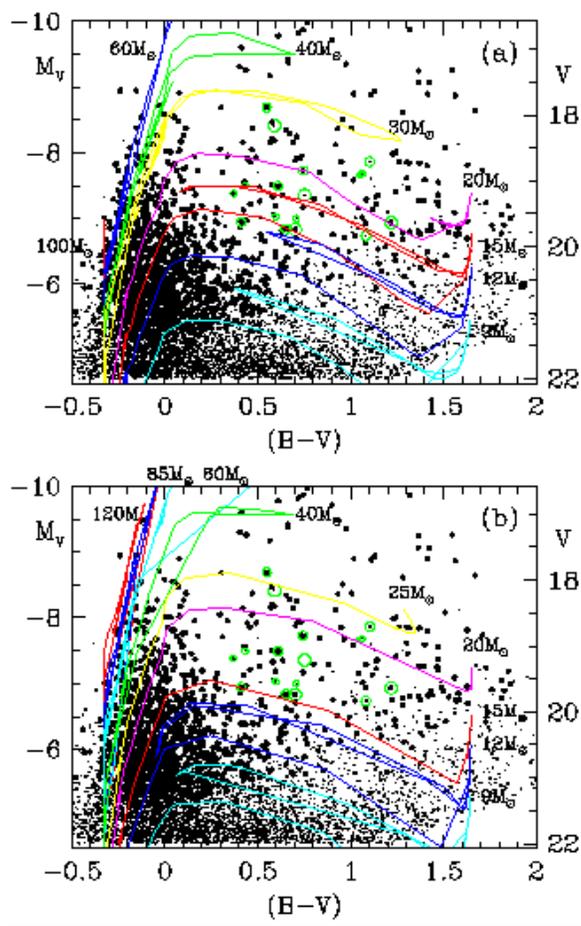}
\figcaption{Observation HR Diagram of NGC 300. (a) The ($M_V,
B-V$) diagram. The superimposed evolution tracks from Bertelli et al. (1994) are
for a metallicity of Z = 0.02. The transformation to the observational
plane was performed using the calibration of Sung (1995). (b) The same as (a),
but the superimposed evolution tracks are from Schaller et al. (1992). All the symbols
are the same as in Fig.~3.
}
\end{figure*}

\begin{figure*}[9]
\plotone{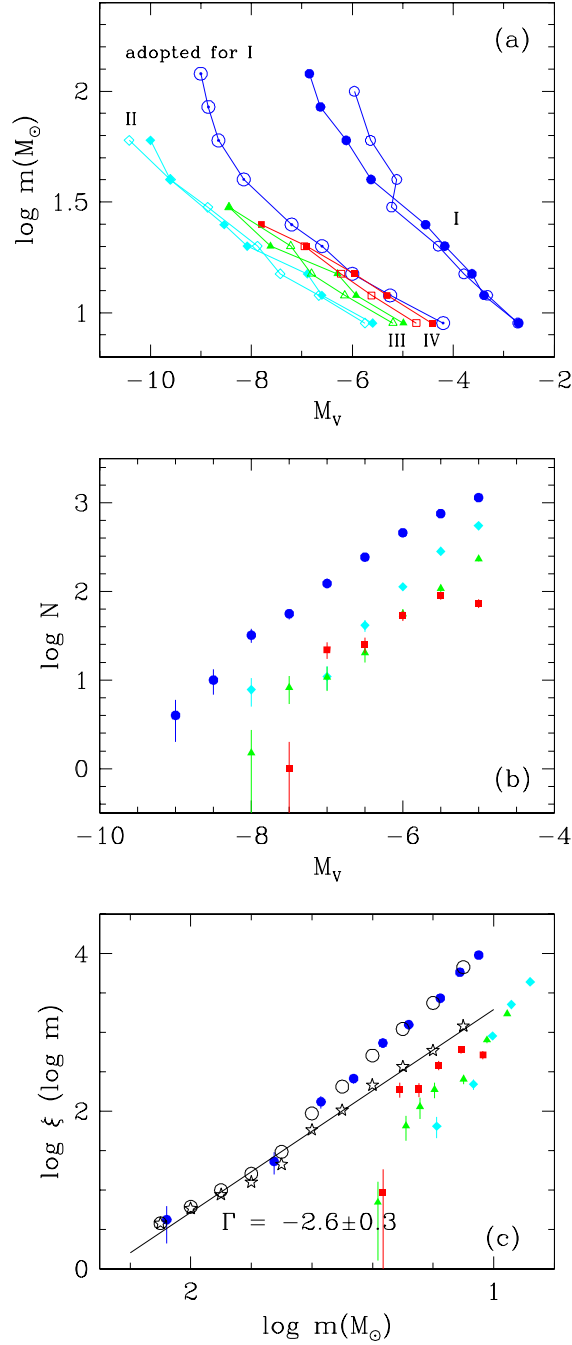}
\figcaption{The luminosity and mass function 
of the bright stars in NGC 300. (a) The
mass - luminosity relation for each group. Open and filled symbols represent,
respectively, the relation from Bertelli et al. (1994) and Schaller et al. (1992). 
The circle,
diamond, triangle, and square denote, respectively, the theoretical mass - luminosity
relations for group I, II, III, and IV. The circle with dot represents 
the adopted relation for group I (blue stars). See text for more details.
(b) The luminosity function. The contribution of foreground field stars
were subtracted using the model calculation by Ratnatunga \& Bahcall (1985). 
(c) The mass
function of NGC 300. The mass function of each group was derived from
the luminosity function obtained in (b) using the mass - luminosity
relation in (a). The open circle and star mark represents, respectively,
the total mass function and the initial mass function of NGC 300. See
text for the derivation of the initial mass function.
}
\end{figure*}

\begin{figure*}[10]
\plotone{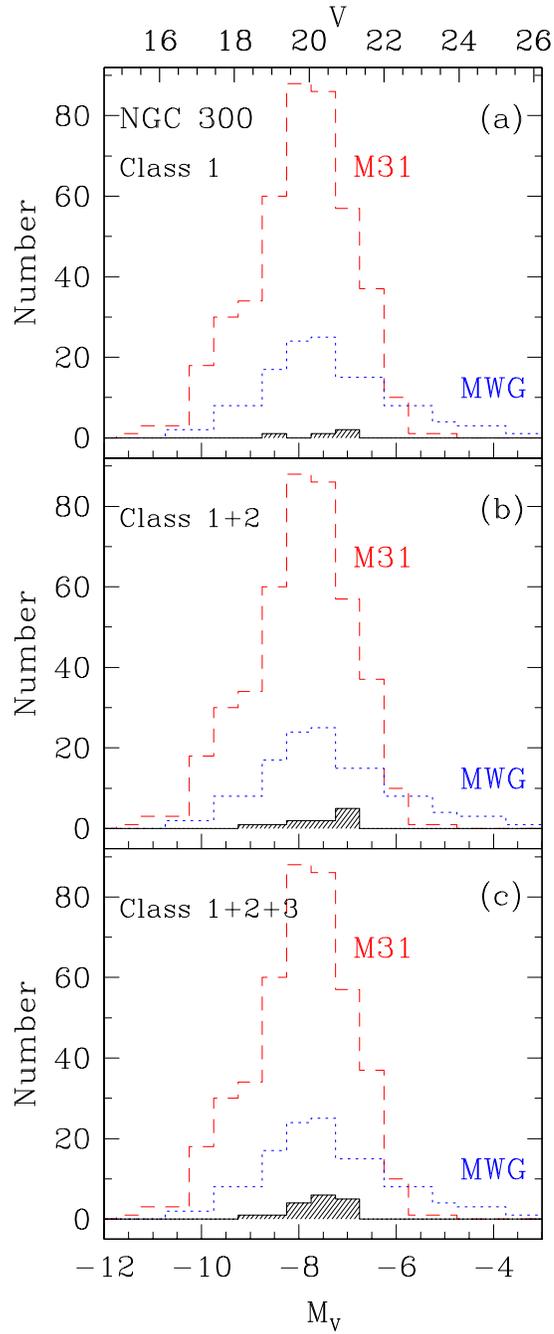}
\figcaption{Luminosity function of the GC candidates 
in NGC 300 (solid lines).
Data for the MWG GCs (Harris 1996; dotted lines) and 
M31 GCs (Barmby et al. 2000; dashed lines) are also plotted for comparison.
}
\end{figure*}

\begin{figure*}[11]
\plotone{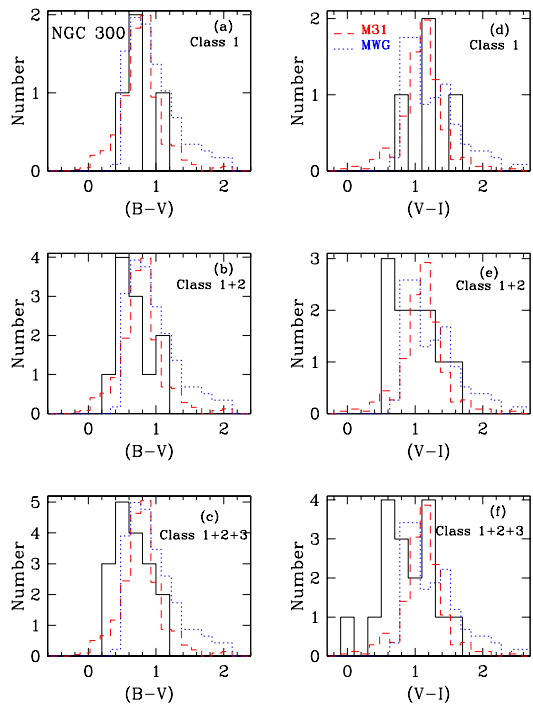}
\figcaption{Color distribution of the GC candidates 
in NGC 300 (solid lines).
Data for the MWG GCs (Harris 1996; dotted lines) and 
M31 GCs (Barmby et al. 2000; dashed lines) are also plotted for comparison.
The labels in the ordinates are for the GC candidates in NGC 300 and
are in arbitrary scales for the GCs in M31 and MWG.
}
\end{figure*}

\begin{figure*}[12]
\plotone{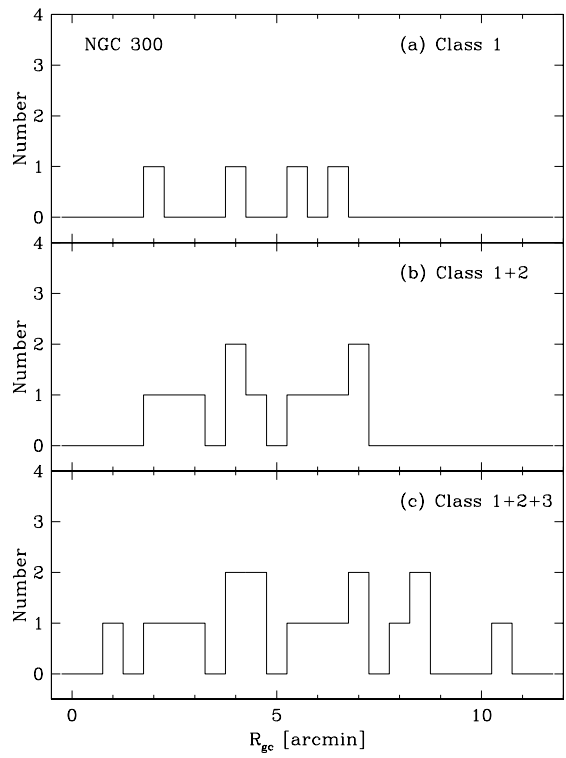}
\figcaption{Histogram of the projected galactocentric distances
of the GC candidates from the nuclues of NGC 300.
}
\end{figure*}
\end{document}